\documentclass[showpacs,showkeys,superscriptaddress,pra]{revtex4}
\usepackage{amsmath}
\usepackage{amssymb}
\usepackage{bbold}
\usepackage{graphicx}
\usepackage{subfigure}
\usepackage{hyperref}
\usepackage{color, soul}
\input{Qcircuit}

\def\bra#1{\mathinner{\langle{#1}|}}
\def\ket#1{\mathinner{|{#1}\rangle}}

\newcommand{\ketbra}[2]{|#1\rangle \langle #2|}

\def\e{\epsilon}

\begin{document}

\title{The robustness of magic state distillation against errors in Clifford gates}

\author{Tomas Jochym-O'Connor}
\email{trjochym@uwaterloo.ca}
\affiliation{
    Institute for Quantum Computing and Department of Physics and Astronomy,
    University of Waterloo,
    200 University Avenue West, Waterloo, Ontario, N2L 3G1, Canada
    }

\author{Yafei Yu}
\affiliation{
    Institute for Quantum Computing and Department of Physics and Astronomy,
    University of Waterloo,
    200 University Avenue West, Waterloo, Ontario, N2L 3G1, Canada
    }
\affiliation{Laboratory of Nanophotonic Functional Materials and Devices, SIPSE \& LQIT, South China Normal University, Guangzhou 510006, China}

\author{Bassam Helou}
\affiliation{
    Institute for Quantum Computing and Department of Physics and Astronomy,
    University of Waterloo,
    200 University Avenue West, Waterloo, Ontario, N2L 3G1, Canada
    }

\author{Raymond Laflamme}
\affiliation{
    Institute for Quantum Computing and Department of Physics and Astronomy,
    University of Waterloo,
    200 University Avenue West, Waterloo, Ontario, N2L 3G1, Canada
    }
\affiliation{
   Perimeter Institute for Theoretical Physics, 31 Caroline Street North, Waterloo
Ontario, N2L 2Y5, Canada}

\begin{abstract}

Quantum error correction and fault-tolerance have provided the possibility for large scale quantum computations without a detrimental loss of quantum information. A very natural class of gates for fault-tolerant quantum computation is the Clifford gate set and as such their usefulness for universal quantum computation is of great interest. Clifford group gates augmented by magic state preparation give the possibility of simulating universal quantum computation. However, experimentally one cannot expect to perfectly prepare magic states. Nonetheless, it has been shown that by repeatedly applying operations from the Clifford group and measurements in the Pauli basis, the fidelity of noisy prepared magic states can be increased arbitrarily close to a pure magic state~\cite{Bravyi}. We investigate the robustness of magic state distillation to perturbations of the initial states to arbitrary locations in the Bloch sphere due to noise. Additionally, we consider a depolarizing noise model on the quantum gates in the decoding section of the distillation protocol and demonstrate its effect on the convergence rate and threshold value. Finally, we establish that faulty magic state distillation is more efficient than fault-tolerance-assisted magic state distillation at low error rates due to the large overhead in the number of quantum gates and qubits required in a fault-tolerance architecture. The ability to perform magic state distillation with noisy gates leads us to conclude that this could be a realistic scheme for future small-scale quantum computing devices as fault-tolerance need only be used in the final steps of the protocol. 
\end{abstract}


\maketitle

%%
%% SECTION
%%
\section{Introduction}
\label{sec:introduction}

Processes such as imperfect control of quantum operations or unintended coupling between qubit systems and their environment lead to errors in any realistic implementation of a quantum computing device. As such, quantum error correction has been developed in order to recover the quantum information that would otherwise be lost due to these faults~\cite{Landauer, ShorEC, SteaneEC}. However, quantum error correction itself is not enough for the implementation of a robust quantum computing device as errors that occur between the error correction steps could propagate between qubits. Propagating errors could prove to be detrimental to the recovery of quantum information and need to be avoided in order to implement any realistic error correction scheme. Fault-tolerant quantum computation aims to address this concern by encoding the information of each qubit into a larger Hilbert space of many qubits and perform encoded quantum gates in such a way that errors do not propagate through multiple qubit blocks~\cite{ShorFT}. Fault-tolerance allows the faults to remain tractable at the cost of needing a polylogarithmic increase in the number of qubits and quantum gates to perform encoded operations when the error rate of the quantum gates is below a certain target threshold~\cite{Gottesman97, KLZ, Preskill, Aharonov, AGP06}.

A desired property of encoded gate operations is transversality, which prevents the propagation of errors in the encoded states of the fault-tolerant architecture. The Clifford gate set, the group of gates generated by the Hadamard gate, the phase gate, and CNOT gate, has been shown to be transversal for many quantum codes~\cite{Nielsenbook}. The Clifford gates, along with measurement and state preparation in the $Z$-eigenbasis form the class of stabilizer operations, which have been shown to be efficiently classically simulatable~\cite{AG04fromBen}. In order to use Clifford gates for universal quantum computation, one requires the ability to produce a pure, single-qubit, non-stabilizer state~\cite{Knill2004}, known as a magic state. Perfect magic state preparation is difficult due to experimental errors, however magic state distillation allows for the creation of an arbitrarily high fidelity magic state from noisy ancillas by repeatedly applying stabilizer operations~\cite{Bravyi, Ben, Ben2}. In this work we investigate the effect of noise, in the state preparation and quantum gate application, on the convergence of the magic state distillation protocol. 

In Bravyi and Kitaev's magic state distillation protocol~\cite{Bravyi}, five copies of a noisy magic state are used to extract a single state of higher fidelity with respect to the magic state. The process is then iterated to increase the fidelity to be arbitrarily high. The input states to the distillation scheme are assumed to be along the magic state axis in the Bloch sphere, the axis connecting the magic state and its orthogonal complement. This assumption is based upon the ability to perform a dephasing channel that collapses all points of the Bloch sphere to their projection along this axis. In this work, we extend the analysis of the distillation scheme to one where the location of the input state is an arbitrary state in the Bloch sphere. The motivation for this is twofold, performing a dephasing operation in an experimental setup would introduce errors that would manifest themselves as perturbations off the magic state axis; moreover, there may be states off the magic state axis that converge faster to the magic state after multiple iterations than the states with the same fidelity on the magic state axis. Performing such an analysis will enable us to conclude that the magic state distillation protocol is robust to slight perturbations about the magic state axis in the Bloch sphere due to noise. 

In Section~\ref{sec:GateNoise}, we turn our attention to the effect of noise present in the quantum gates of the distillation protocol. We investigate the consequences of depolarizing noise, a one-parameter noise model, on the one and two-qubit Clifford gates in the five-qubit decoder of the protocol. Such noise will affect the rate of convergence, increase the threshold for the input states of the protocol, and pose a restriction on the ability to prepare a magic state with arbitrarily high fidelity. As such, with the development of fault-tolerant schemes for the reduction of noise, one can ask if it is more efficient in the total number of quantum gates to use faulty gates or fault-tolerant encoded gates to perform the distillation protocol. In Section~\ref{sec:GateComparison} we show that using faulty Clifford gates is the more efficient scheme and using a fault-tolerant encoding is unnecessary unless the required gate fidelity of the universal gate is much higher than that of the Clifford gates at one's disposal. The ability to perform magic state distillation without the use of a fault-tolerant encoding is promising for future experimental realizations of multiple rounds of state distillation on small-scale quantum computers. 

%%
%% SECTION
%%
\section{The evolution of quantum states under magic state distillation with perfect Clifford gates}
\label{sec:nodephasing}

The input states to Bravyi and Kitaev's magic state distillation protocol are assumed to be along the magic state axis~\cite{Bravyi}. However, preparing such states may prove to be difficult experimentally. In this section we study the convergence of the distillation scheme under perturbations about the magic state axis. Moreover, we show that for low fidelity input states, the convergence to the magic state may be improved for states away from the magic state axis. This suggests that while performing a dephasing operation to initialize the input states to be along the axis may be useful, it is not absolutely necessary in certain fidelity regimes. 

The Clifford gate set is generated by the following gates,
\begin{align}
H = \dfrac{1}{\sqrt{2}} \begin{pmatrix} 1 & 1\\ 1 & -1 \end{pmatrix}, \qquad K = \begin{pmatrix} 1 & 0\\ 0 & i \end{pmatrix}, \qquad CNOT = \begin{pmatrix} I & 0\\ 0 & \sigma^x \end{pmatrix},
\end{align}
where each individual gate can be performed on any qubit, and CNOT can be performed on any pair of qubits where $I$ and $\sigma^x$ are 2--by--2~Pauli matricies. The power of magic state distillation is that it requires only the use of Clifford gate operations, along with $\ket{0}$ state ancilla preperation and measurement in the $Z$-basis, to distill multiple copies of noisy magic states to one of higher fidelity, provided the initial state is above a given threshold. This is appealing as Clifford gates have been shown to have transversality in many quantum error correcting codes~\cite{Nielsenbook}, and as such can be implemented fault-tolerantly in order to reduce their error rate.

Let $\sigma^i$ denote the $i^{\text{th}}$~Pauli matrix in the computational basis. There are two types of magic states, up to one-qubit Clifford operators, the H-type magic state,
\begin{align*}
\ketbra{H}{H}= \frac{1}{2}\left[I+\frac{1}{\sqrt{2}}(\sigma^{x}+\sigma^{z})\right],
\end{align*}
 which can be used to implement the $\pi/8$-phase gate, $\Lambda_{\pi/8}~=~e^{i\pi/8}\ketbra{0}{0} + e^{-i\pi/8} \ketbra{1}{1}$, and the T-type magic state,
\begin{align*}
\ketbra{T}{T} = \frac{1}{2}\left[I+\frac{1}{\sqrt{3}}(\sigma^{x}+\sigma^{y}+\sigma^{z})\right],
\end{align*}
which can be used to implement the $\pi/12$-phase gate $\Lambda_{\pi/12}$~\cite{Bravyi}, both of which, along with the Clifford gates, provide universal quantum computation. Many efforts have contributed to building protocols for magic state distillation and achieving tight noise thresholds for the noisy input ancillas to the distillation protocol in order to understand
the transition from classically simulatable quantum computation to genuine quantum computation~\cite{Bravyi,Ben, Ben2,Campbell1, Campbell2, Anwar, Veitch, Anderson}. Additionally, an experimental demonstration of a single round of Bravyi and Kitaev's distillation protocol of T-type magic states has been performed in Nuclear Magnetic Resonance (NMR)~\cite{jingfu}.

The first step of Bravyi and Kitaev's distillation protocol~\cite{Bravyi} is to perform a dephasing operation $\mathcal{T}$ on five copies of the initial state of the quantum system,
\begin{align}
\mathcal{T}(\rho) = \dfrac{1}{3} \left( \rho + T \rho T^{\dagger} + T^{\dagger} \rho T \right),
\end{align}
where $T=KH$ is a Clifford group gate. If the initial state of the system is expressed according to its Bloch sphere coordinates~$(x,y,z)$,
\begin{align}
\rho=\frac{1}{2}\left[I+x\sigma^{x}+y\sigma^{y}+z\sigma^{z}\right],
\end{align}
the transfomation $\mathcal{T}$ is equivalent to projecting the state $(x,y,z)$ onto the magic state axis connecting the states~$\ket{T_0}$ and $\ket{T_1}$~in the Bloch sphere, where $\ket{T_1}$ is the state orthogonal to $\ket{T_0}$,
\begin{align}
\mathcal{T}(\rho)=\frac{1}{2}\left[I+\dfrac{x+y+z}{3}(\sigma^{x}+\sigma^{y}+\sigma^{z})\right].
\end{align}
The dephasing operation leads to the ability to derive a clean threshold for the input fidelity of the initial states in order for the magic state distillation protocol to be beneficial. However, errors in the implementation of the quantum information processor could lead to a preparation of states away from the magic state axis. In this section, we provide an analysis of the effectiveness of the magic state distillation protocol for states prepared at an arbitrary location in the Bloch sphere and give modified target fidelities for state distillation under such noisy state preparation.

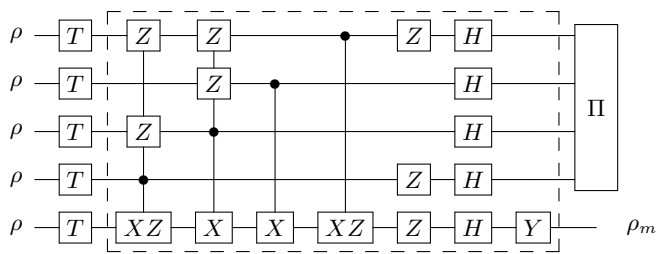
\begin{figure}[h]
\begin{align*}
\Qcircuit @C=1em @R=.7em {
	\lstick{\rho}	& \gate{T}	& \gate{Z}				& \gate{Z}				& \qw	& \ctrl{4}	& \gate{Z}	& \gate{H}	& \qw	& \multigate{3}{ \Pi} &\\
	\lstick{\rho}	& \gate{T}	& \qw				& \gate{Z}	\qwx[-1]		& \ctrl{3}	& \qw	& \qw	& \gate{H}	& \qw	& \ghost{\Pi}		&\\
	\lstick{\rho}	& \gate{T}	& \gate{Z}	\qwx[-2]		& \ctrl{2} \qwx[-1] 		& \qw	& \qw	& \qw	& \gate{H}	& \qw	& \ghost {\Pi}		&\\
	\lstick{\rho}	& \gate{T} & \ctrl{1} \qwx[-1] 		& \qw				& \qw	&\qw		& \gate{Z}	& \gate{H}	& \qw	& \ghost{\Pi} 		&\\
	\lstick{\rho}	& \gate{T}	& \gate{XZ}			& \gate{X}				& \gate{X}	&\gate{XZ}& \gate{Z}	& \gate{H}	& \gate{Y} & \qw			& \rho_{m} \gategroup{1}{3}{5}{9}{.7em}{--}
}
\end{align*}
\caption{Analyzed circuit for the five-qubit magic state distillation protocol. The gate $\mathcal{T}$ represents the dephasing transformation. The gates $X$ and $Z$ are the standard Pauli gates, and the gate $H$ is the Hadamard gate. The symbol $\Pi$ represents the projection onto the trivial subspace upon post-selecting the measurement outcome ``+1" for all the $Z$-basis measurements. All gates contained in the boxed region represent the decoding circuit for the five-qubit error correcting code. }
\label{fig:perfect_circuit}
\end{figure}

The distillation protocol consists of the above dephasing transformation on five prepared initial states, followed by a measurement of the stabilizers of the five-qubit code~\cite{Bennett96, Laflamme96}, $XZZXI$, $IXZZX$, $XIXZZ$, $ZXIXZ$, and decoding upon obtaining the~``+1"~eigenstate of all the stabilizers~\cite{Bravyi}, where $X=\sigma^x$ and $Z=\sigma^z$. The roles of the measurement and decoding can be reversed, and the overall circuit can be described by the diagram in Figure~\ref{fig:perfect_circuit}. Various encoding/decoding circuits can be developed to encode the five qubit code, we chose to analyze the circuit presented in Figure~\ref{fig:perfect_circuit} as once a qubit is used as a control qubit in a two-qubit gate, it is no longer used in any further two-qubit gates, thus minimizing the propagation of errors through the circuit~\cite{Grassl}. Upon following the steps outlined in the above procedure, the initial noisy states must have a fidelity greater than~$F_{T}=\frac{1}{2}(1+\sqrt{\frac{3}{7}})\approx0.8273$ with respect to the magic state in order for the output state to be of higher fidelity~\cite{Bravyi}, where the fidelity with respect to the magic state is defined as~$F(\rho)=\bra{T_0} \rho \ket{T_0}$. Repeating the protocol to obtain multiple copies of the state of increased fidelity $\rho_m$, the process can be iterated to obtain magic states with arbitrarily high fidelity.

% CIRCUIT 1 (Magic state decoding circuit)

We shall consider the scenario where the dephasing gates are omitted from the protocol, and only the gates in the dashed box in Figure~\ref{fig:perfect_circuit} are implemented, followed by post-selection upon obtaining outcomes of~``+1" for $Z$-basis measurements on the top four qubits. The input state is now located at an arbitrary position in the Bloch sphere given by coordinates~$(x,y,z)$, then the output state coordinates are as follows:
\begin{align}
	x_{out}=\dfrac{-z(z^{4}-5x^{2}+5y^{2}(x^{2}-1))}{(1+5z^{2}y^{2}+ 5x^{2}y^{2}+5z^{2}x^{2})}. \nonumber\\
        	y_{out}=\dfrac{-y(y^{4}-5z^{2}+5x^{2}(z^{2}-1))}{(1+5z^{2}y^{2}+ 5x^{2}y^{2}+5z^{2}x^{2})},\\
        	z_{out}=\dfrac{-x(x^{4}-5z^{2}+5y^{2}(z^{2}-1))}{(1+5z^{2}y^{2}+ 5x^{2}y^{2}+5z^{2}x^{2})},\nonumber 
\end{align}

\begin{figure}[h]
\includegraphics[width=0.55\textwidth]{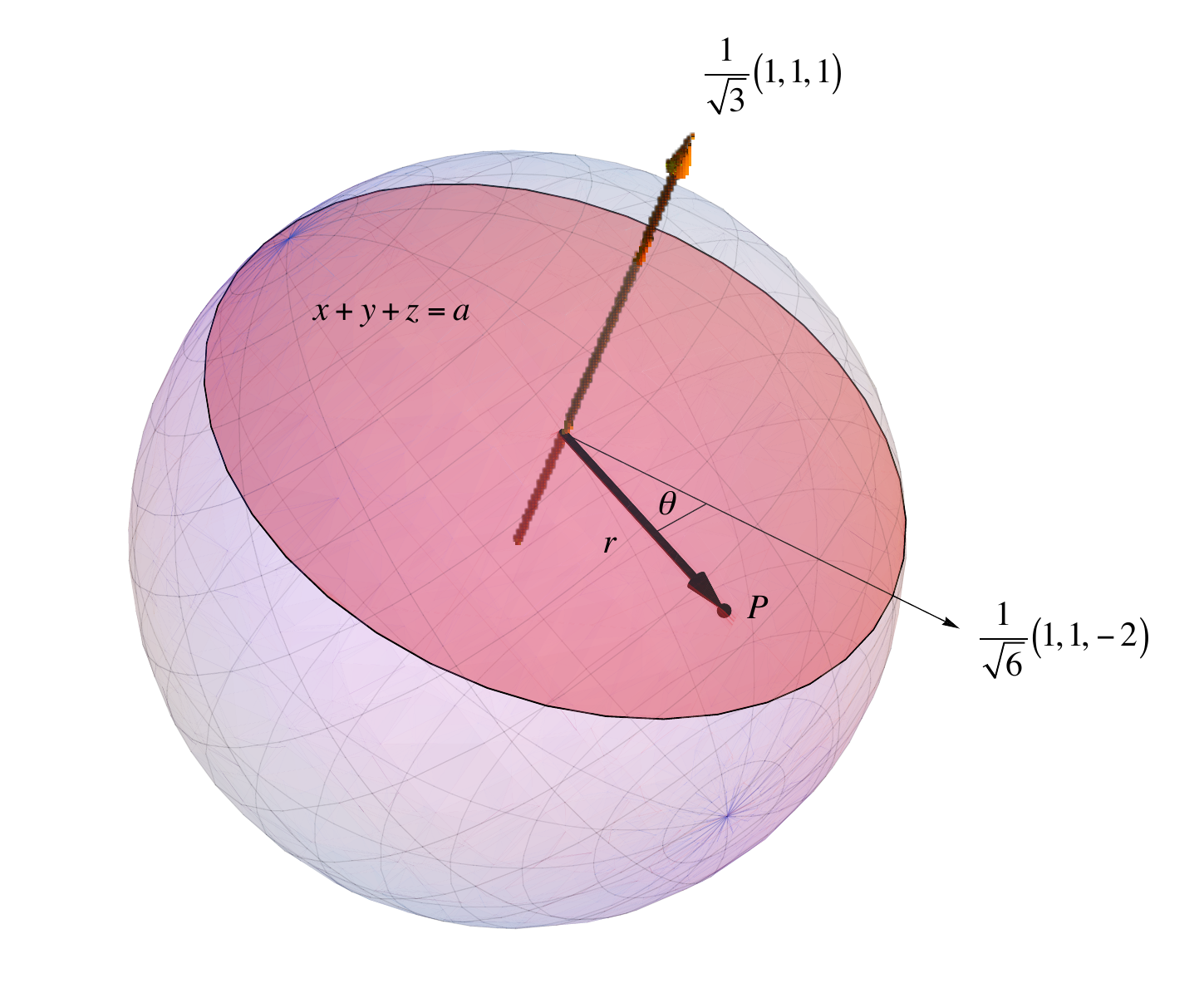}
\caption{Labelling of the coordinates in the Bloch sphere used to derived the fidelity difference after a single run of the magic state distillation protocol, given in Eq.~\ref{eq:fidelity_increase}. The red plane is the plane of constant fidelity~$x+y+z=a$, the orange arrow is the magic state axis, labeled by its normalized Bloch~sphere coordinates~$(1,1,1)/\sqrt{3}$, and the axis labelled by the vector~$(1,1,-2)/\sqrt{6}$ is the modified $x$-axis. For a given point~$P$, the coordinates $a$,~$r$,~and~$\theta$ are defined as in the figure.}
\label{fig:labels}
\end{figure}

The plane of states that have fidelity $F$ are the states in the Bloch sphere that satisfy,
\begin{align}
\label{eq:fidelity_plane}
x+y+z = \sqrt{3}(2F-1).
\end{align}
For a given plane of constant input fidelity $F_{in}$, define a new coordinate system for that plane where $r$ is the radial distance of the input state from the magic state axis, and $\theta$ as the angle between the distance vector and the modified $x$~axis, as shown in Figure~\ref{fig:labels}. The difference between the input fidelity and output fidelity can then be expressed according to these coordinates as
\begin{align}
\label{eq:fidelity_increase}
    d = F_{out}-F_{in}=-\dfrac{2\Big( a(54-60a^2+14a^4+135r^4)+15\sqrt{6}(-3+a^2)r^3\cos{3\theta} \Big)}{2\sqrt{3}\Big( 108+20a^4+135r^4 +60 \sqrt{6} a r^3 \cos{3\theta} \Big) } ,
\end{align}
%\end{widetext}
where $a$ is related to the input fidelity, $a=\sqrt{3}(2F_{in}-1)$. As such, the ability of the distillation protocol to increase the fidelity of the input states depends on the distance of the initial states from the magic state axis and on the spatial angle with respect to the modified $x$~axis as well as the input fidelity.

\begin{figure}[h]
\includegraphics[trim=2cm 13cm 1.5cm 0cm, clip, width=0.4\textwidth]{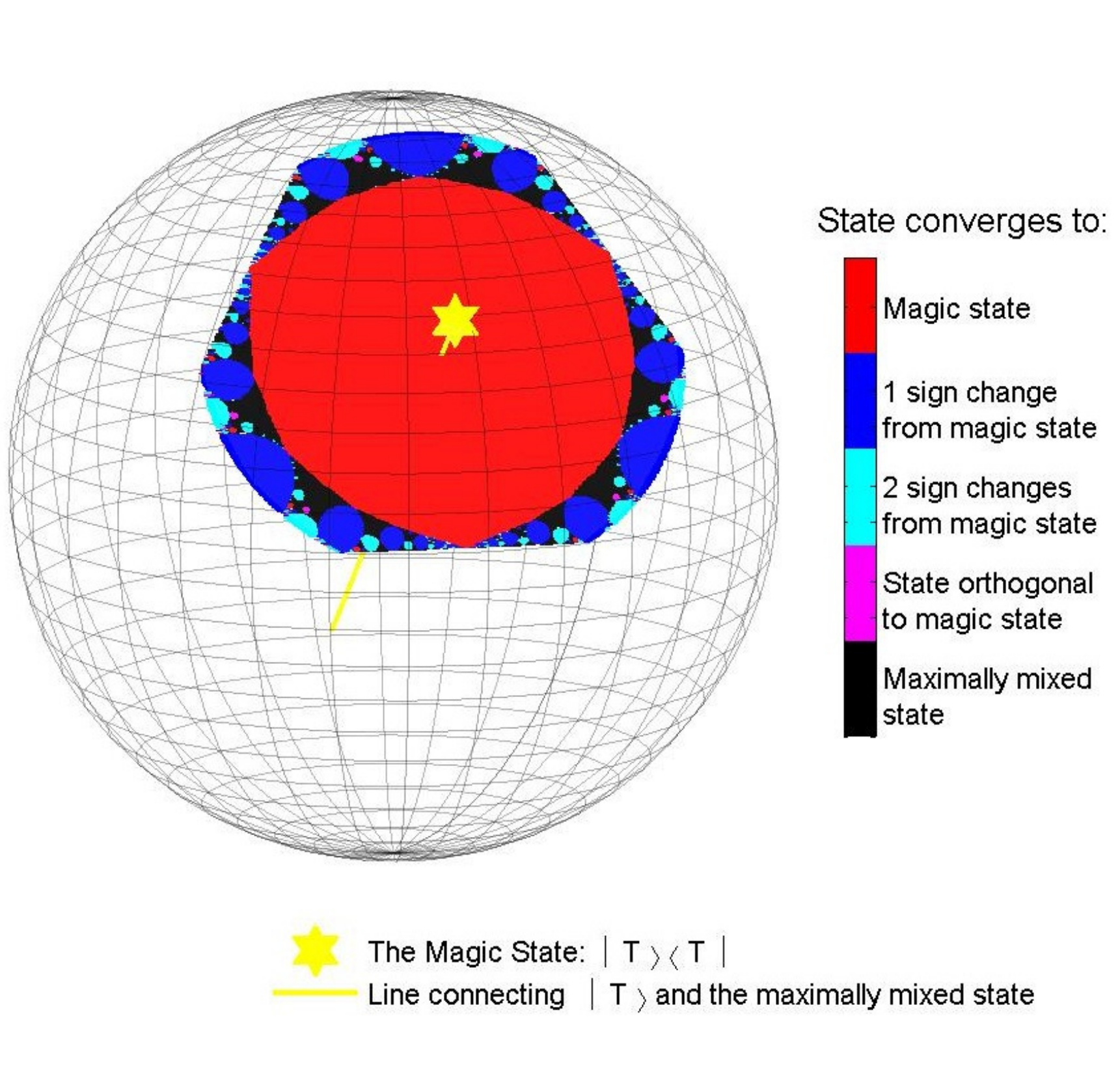}
\caption{Final states of convergence after multiple rounds of distillation for states on the fidelity plane $F=0.886$, that is a cut of the Bloch sphere with states of a fixed fidelity, as given by Eq.~\ref{eq:fidelity_plane}. The states in red represent the states that converge to the magic state after multiple rounds of distillation. The states in light and dark blue converge to states with the coordinates $(\pm1,\pm1,\pm1)/\sqrt{3}$ in the Bloch sphere, where the number of sign changes from the coordinates of the $\ket{T}$ magic state $(1,1,1)/\sqrt{3}$ is given by the shading of blue. The states in pink converge to the state orthogonal to $\ket{T}$, and black to the maximally mixed state.}
\label{fig:0.886fid}
\end{figure}

\begin{figure}[h]
%\subfigure{\includegraphics[width=0.45\textwidth]{FIgures/082a}}
%\subfigure{\includegraphics[width=0.35\textwidth]{FIgures/082b}}
%\includegraphics[trim=2.5cm 22cm 2cm 0cm, clip, width=0.9\textwidth]{Figures/082.eps}
\includegraphics[trim=2.5cm 22cm 2cm 0cm, clip, width=0.9\textwidth]{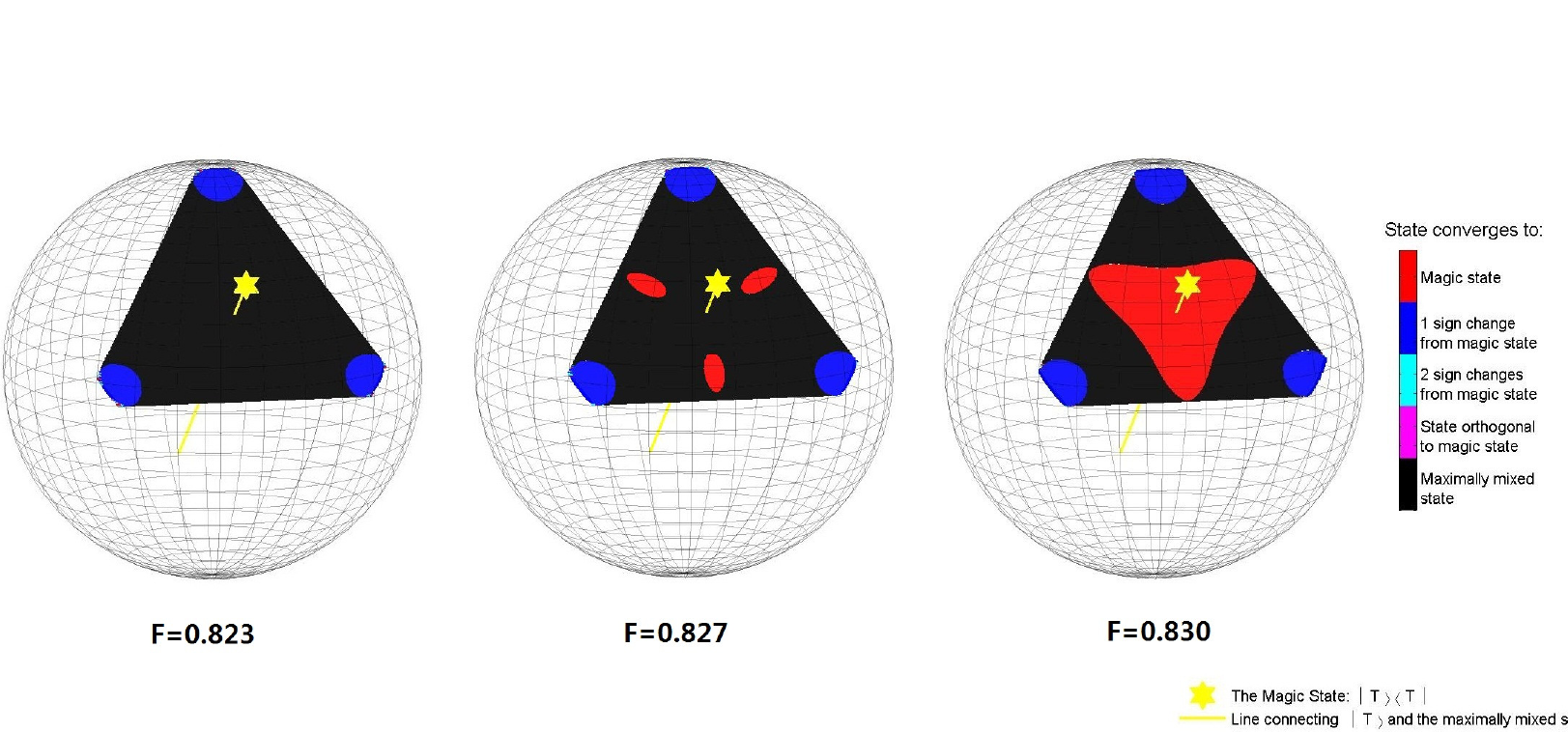}
\caption{Convergence of the states around the perfect Clifford gate threshold 0.8273 for states along the magic state axis. All states much below the theoretical threshold ($F=0.823$) converge away from the magic state axis, yet some states slightly below the threshold for on axis convergence converge to the magic state ($F=0.8270$). This convergence only appears for states close to the three angles of maximum output fidelity on the fidelity plane. All states close to the magic state axis converge to the magic state for fidelities above the threshold ($F=0.830$).}
\label{fig:0.827fid}
\end{figure}

\begin{figure}[h]
%\subfigure{\includegraphics[width=0.4\textwidth]{Figures/distance1.eps} \label{fig:maximal_theta}}
\subfigure{\includegraphics[width=0.4\textwidth]{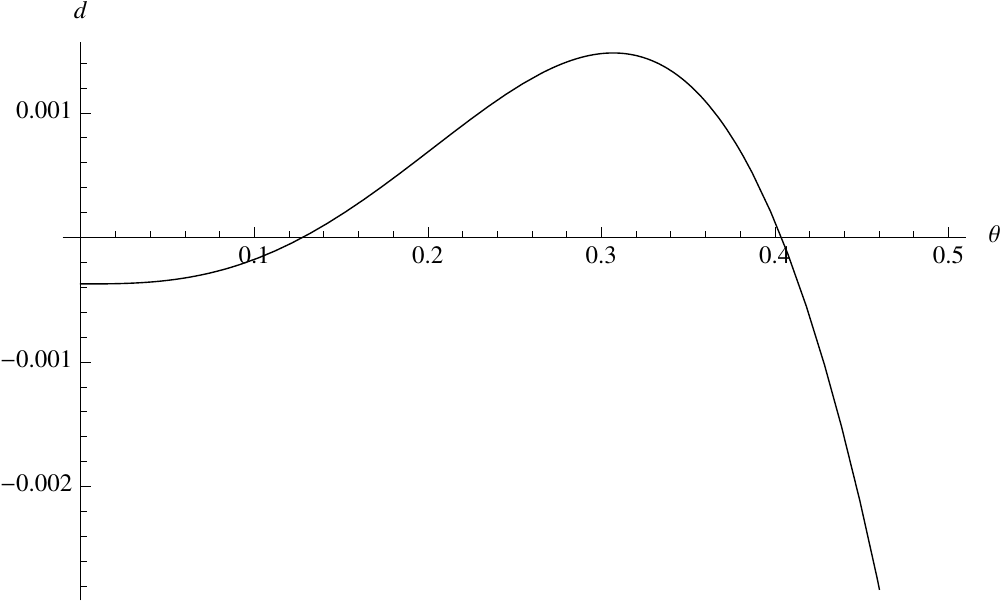} \label{fig:maximal_theta}}
\hspace{1cm}
%\subfigure{\includegraphics[width=0.4\textwidth]{Figures/angle.eps} \label{fig:angular_dependance}}
\subfigure{\includegraphics[width=0.4\textwidth]{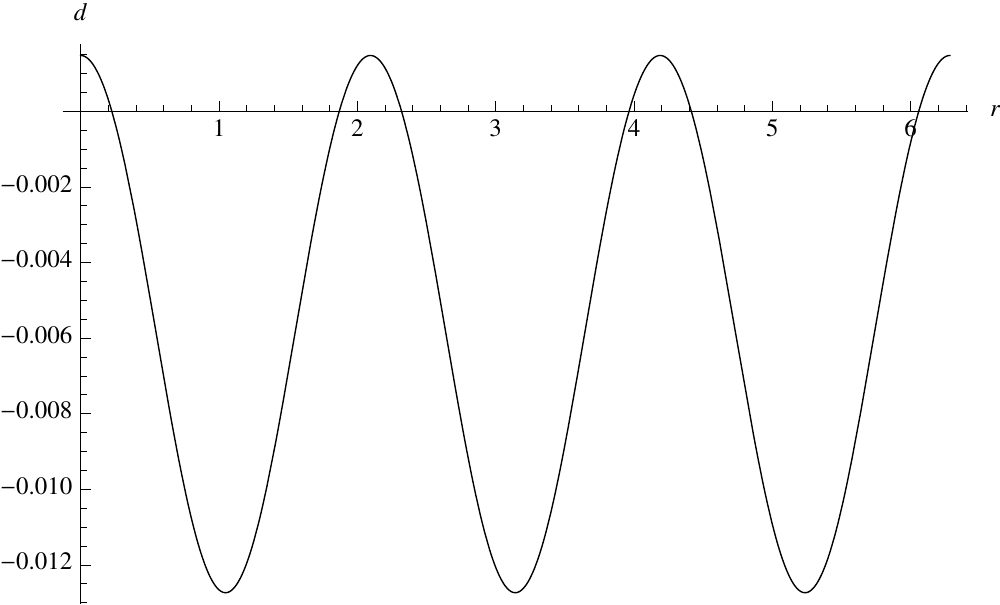} \label{fig:angular_dependance}}
\caption{(a) Dependance of the difference in fidelity $d$, as defined in Equation~\eqref{eq:fidelity_increase}, between the initial and final states as a function of distance from the magic state axis in the Bloch sphere, at fixed angle $\theta = 0$ and fidelity $F=0.8269$. (b) Dependance of distance in fidelity between the initial and final states as a function of angle for fixed distance from the Bloch sphere $r = 0.3$ and fidelity $F=0.8269$.}
\end{figure}

Figure~\ref{fig:0.886fid} demonstrates the dependence on the distance from the magic state axis and angle for convergence to the magic state after many iterations for the set of states on the plane with fidelity~$F=0.886$ in the positive octant~($x,y,z>0$). States close to the magic state axis converge to the magic state, while states far away from the axis converge  to other undesirable states in the Bloch sphere. Perhaps more surprisingly, there are states below the distillation threshold set by Bravyi and Kitaev~\cite{Bravyi} that can converge to the magic state. As Figure~\ref{fig:0.827fid} shows, states away from the magic state axis on the fidelity plane of $F=0.8270$ converge to the magic state, while those on the axis converge to the maximally mixed state, as expected as $F$ is below the fidelity threshold for states on the axis. Notice in Eq.~\ref{eq:fidelity_increase} that the maximal increase in the fidelity of the state after the distillation procedure occurs for angles $\theta = 0, \text{ }2\pi/3, \text{ }4\pi/3$. Fixing~$\theta = 0$, the difference in fidelity can show an increase as a function of the distance from the magic state axis~$r$ for small distances before dropping off rapidly as the distance from the axis grows. For an input fidelity just below the fidelity threshold set in Ref.~\cite{Bravyi}, the fidelity difference can be negative for states close to the axis and increase to be positive for certain distances away from the axis, as Figure~\ref{fig:maximal_theta} shows, but only for states whose angle is close to the angles of maximal increase, as plotted in Figure~\ref{fig:angular_dependance}. Therefore at angles close to the angles of maximal increase, one can obtain states whose fidelity threshold can be below the on-axis threshold of~$\frac{1}{2}(1+\sqrt{\frac{3}{7}}) \approx 0.8273$. As such, by performing the magic state distillation routine without the dephasing transformation, one can slightly lower the threshold for the fidelity of input states to be~$0.825$ for states at the angle of maximal increase.

We may thus conclude that the dephasing operation that projects the noisy magic states onto the magic state axis is not necessarily needed as there are regions of convergence around this axis. As such, noisy dephasing processes that would have slight perturbations off the magic state axis would not be detrimental to the convergence to the magic state. Furthermore, for certain states off the magic state axis and of fidelity just below that derived by Bravyi and Kitaev, omitting the dephasing operation would be beneficial as it would allow such states to remain useful for magic state distillation.

%%
%% SECTION
%%
\section{Magic state distillation with noisy Clifford gates}
\label{sec:GateNoise}
\subsection{Error rate of faulty magic state distillation}
\label{sec:depolarizing_noise}

In an experimental realization of the implementation of magic state distillation, any quantum gate will introduce noise. As such, an interesting experimental question would be: to what level of noise is the distillation still beneficial? In order to address such a question, we consider the case where all the gates in the decoding circuit, shown by the boxed region in Figure~\ref{fig:perfect_circuit}, are subjected to depolarizing noise. Depolarizing noise is a common noise model for physical implementations of quantum information processing. A noisy one-qubit Clifford gate $G$ will be modelled as follows:

% CIRCUIT 2 (One-qubit Error)
%\begin{figure}[h]
%\includegraphics[width=0.25\textwidth]{Figures/OneQubit_Noise}
%\end{figure}
\begin{align*}
\Qcircuit @C=1em @R=.7em {
	& \gate{G} & \qw
}
\quad \rightarrow \quad
\Qcircuit @C=1em @R=.7em {
	& \gate{G} & \gate{\Lambda_1} & \qw
},
\end{align*}
where the gate $\Lambda_1$ performs the depolarizing transformation with a noise parameter $p_1$,
\begin{align}
\label{eq:1qubit_depolarizing}
\Lambda_1(\rho, p_1) &\rightarrow (1-p_1) \rho + p_1 \dfrac{I}{2}= (1-\dfrac{3p_1}{4}) \rho + \dfrac{p_1}{4}( \sigma^{x}\rho \sigma^{x} + \sigma^{y} \rho \sigma^{y} + \sigma^{z}\rho \sigma^{z}).
\end{align}
Similarly, we introduce noise to a two-qubit controlled Clifford gate by adding a two-qubit depolarizing gate,
% CIRCUIT 3: Two-qubit Error
%\begin{figure}[h]
%\includegraphics[width=0.25\textwidth]{Figures/TwoQubit_Noise}
%\end{figure}
%\vspace{-0.5cm}
\begin{align*}
\Qcircuit @C=1em @R=1.35em {
	& \ctrl{1} 	& \qw \\
	& \gate{G}	& \qw \\
}
\quad \rightarrow \quad
\Qcircuit @C=1em @R=1em {
	& \ctrl{1} 	& \multigate{1}{\Lambda_2} & \qw \\
	& \gate{G}	& \ghost{\Lambda_2}& \qw \\
},
\end{align*}
where the two-qubit depolarizing gate $\Lambda_2$ is defined by the transformation with a two-qubit noise parameter $p_2$,
\begin{align}
\label{eq:2qubit_depolarizing}
\Lambda_2 (\rho, p_2)& \rightarrow  (1-p_2)\rho + p_2 \frac{I_2}{4} \nonumber \\
&= \left(1-\dfrac{15p_2}{16}\right) \rho + \dfrac{p_2}{16} \Big( (I \otimes \sigma_x) \rho (I \otimes \sigma_x) + (I \otimes \sigma_y) \rho (I \otimes \sigma_y) +\hdots + (\sigma_z \otimes \sigma_y )\rho (\sigma_z \otimes \sigma_y) + (\sigma_z \otimes \sigma_z )\rho (\sigma_z \otimes \sigma_z) \Big).
\end{align}

We shall only consider errors affecting the gates in the decoding procedure. In order to compare the fidelity threshold for the noisy decoding procedure with the ideal magic state distillation protocol proposed by Bravyi and Kitaev~\cite{Bravyi}, we assume the input states to the decoding procedure are along the magic state axis. This assumption is valid in the error regime we will be considering, as depolarizing errors in the dephasing transformation would lead to a negligible deviation of the input states from the magic state axis. We can relate the parameter~$p_{i}$ to the error per gate, which is defined as $E_{i} = 1-F_{i}$, where $F_{i}$ is the fidelity of the gate~$i$. These values are related by the simple relationships: $p_1 = 2E_1$ and $p_2 = 4E_2/3$. 

Assuming the error rate of the one and two-qubit gates are low (omitting quadratic and higher order terms in $p_1$ and $p_2$), the output (unnormalized) state will have the following matrix entries in the $\ket{T_0}$--$\ket{T_1}$ basis:
\begin{align}
\ketbra{T_0}{T_0} &: \dfrac{1 -5p_1 - 8p_2}{6} \Big[ (1-\e)^5 + 5 \e^3 (1-\e)^2 \Big] + \dfrac{p_1}{36} \Big[ 19 -87\e +197\e^2 - 164\e^3 +6\e^4 +32\e^5 \Big] \nonumber \\
& \qquad \qquad \qquad \qquad \qquad \qquad \qquad \qquad \qquad \qquad +\dfrac{p_2}{54} \Big[20 - 44\e +107\e^2 -106\e^3 +28\e^4 +8\e^5 \Big], \\
\ketbra{T_1}{T_1} &: \dfrac{1- 5p_1- 8p_2}{6} \Big[ \e^5 + 5 \e^2 (1-\e)^3 \Big] + \dfrac{p_1}{36} \Big[ 3 +\e +61\e^2 - 180\e^3 +166\e^4 -32\e^5 \Big] \nonumber \\
& \qquad \qquad \qquad \qquad \qquad \qquad \qquad \qquad \qquad \qquad +\dfrac{p_2}{54} \Big[13 - 4\e +37\e^2 -86\e^3 +68\e^4 -8\e^5 \Big], \\
\ketbra{T_0}{T_1}& :\dfrac{p_2(1+i)(-1+2\e)}{432}\Big[(-2+3i)-(1+5i)\sqrt{3}+((-2-9i)+(3+10i)\sqrt{3})\e  \nonumber \\
& \qquad \qquad \qquad \qquad \qquad \qquad +((6+9i)-(3+6i)\sqrt{3})\e^2 + ((-8-6i)+(2-8i)\sqrt{3})\e^3 + (4+4i\sqrt{3})\e^4 \Big],
\end{align}
where $\e = 1 - \bra{T_0} \rho_{\text{in}} \ket{T_0}$ is the error of the initial state. Note that depolarizing noise in the decoding procedure will introduce off diagonal terms in the $\ket{T_0}$--$\ket{T_1}$ basis, that is, will produce an output state that deviates away from the magic state axis. However, such matrix elements will be on the order of $p_2$, which for low levels of noise will be negligible compared to the strength of the coefficients along the diagonal terms. Therefore, we may assume that upon iterating the distillation protocol, the input states will always remain along the magic state axis. This assumption is also well motivated from the results of Section~\ref{sec:nodephasing} that show that slight deviations off the magic state axis will not affect the convergence of the distillation scheme. For small values of $\e$, the output error $\e_{\text{out}}$, the normalized coefficient of the $\ketbra{T_1}{T_1}$ term, can be approximated to be $5\e^2 + p_1/2 + 13p_2/9$. Thus in the limit of infinite iterations of the distillation protocol the error rate will be linear in the terms $p_1$ and $p_2$. 

\begin{figure}[h!]
\begin{center}
\includegraphics[width = 0.50 \textwidth]{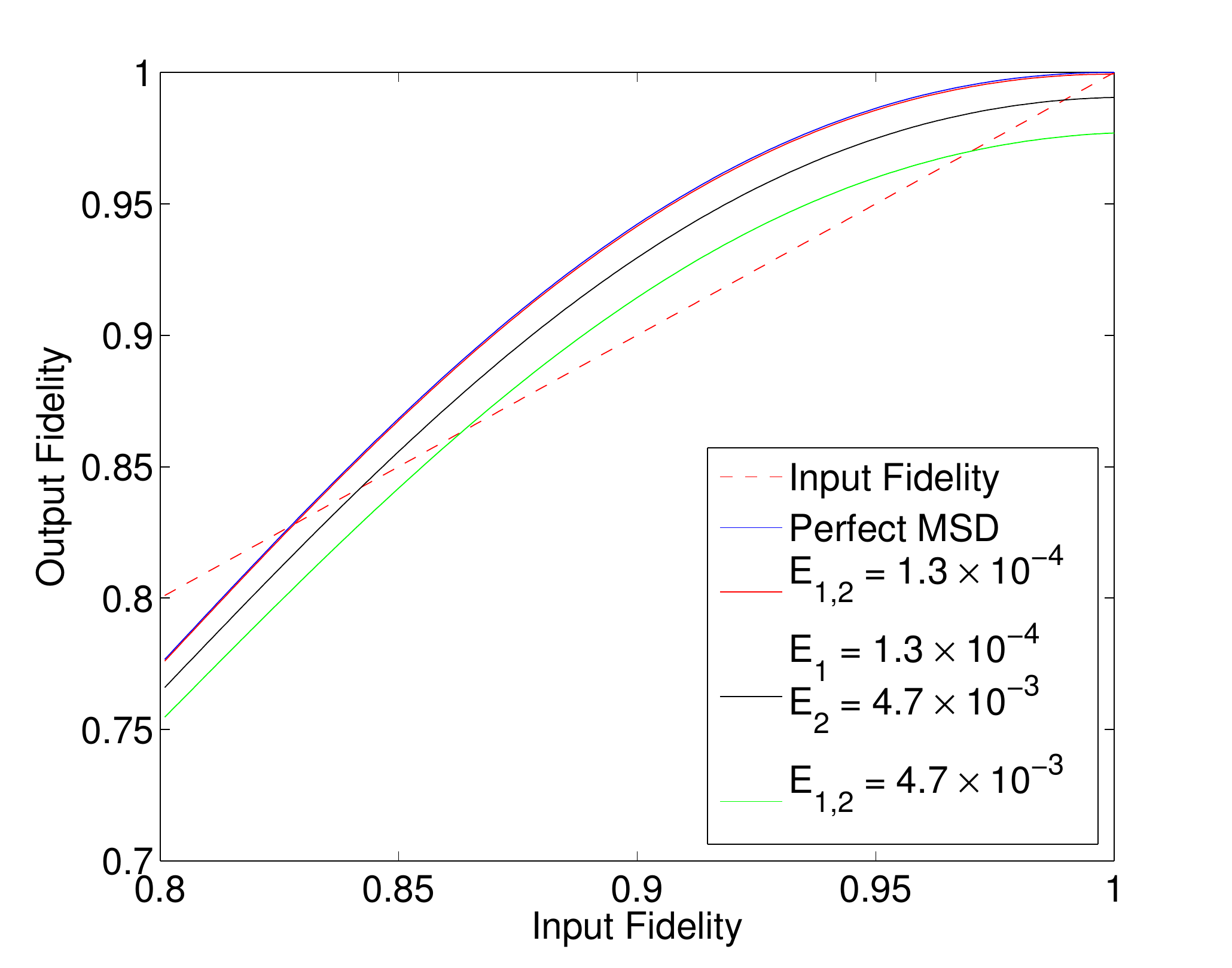}
\caption{Output fidelities of the magic state distillation protocol as a function of the input fidelity of the five input states. The uppermost line indicates the noiseless magic state distillation, while the three other lines show a decrease of the output fidelity caused by an increase in the values of $E_1$ and $E_2$. As the error strength increases, the class of states for which the distillation protocol is beneficial decreases. Notice that once errors are introduced one can no longer increase the fidelity to be arbitrarily close to 1 for repeated applications of the protocol. 
}
\label{fig:MSD_faulty}
\end{center}
\end{figure}

Motivated by the average one and two-qubit gate errors in a recent benchmarking experiment in nuclear NMR~\cite{Ryan_benchmarking}, we have chosen different values for the noise parameters $p_1 $ and $p_2$. The graph in Figure~\ref{fig:MSD_faulty} shows the output fidelity of the magic state distillation protocol for different error strengths of the one and two-qubit gates. The results for error strengths given by the benchmarking NMR results, $E_1 = 1.3 \times 10^{-4}$ and $E_2 = 4.7 \times 10^{-3}$, are given by the black curve. In order to obtain an appreciation of the decrease in output fidelity cause by the strength of both of these errors, we have also plotted the output fidelity corresponding to the case when the one and two-qubit errors have the same strength, namely $1.3 \times 10^{-4}$ and $4.7 \times 10^{-3}$. Note that when both errors are on the order of $10^{-4}$ the decrease in fidelity is minimal, this leads one to conclude that the decrease in fidelity in the black curve is caused mostly by the larger error on the two-qubit gates. Finally, note that for the case of choosing an error model based on the results from Ref.~\cite{Ryan_benchmarking}, the new threshold for the minimal input fidelity is 0.842, which is larger than the theoretical noiseless threshold of 0.8273, and the maximum output fidelity that can be reached through repeated applications of the distillation protocol is 0.9895 with respect to the magic state.

% CIRCUIT 5: Universal Gate Implementation
\begin{figure}[h]
\begin{align*}
\Qcircuit @C=1em @R=1em {
	\lstick{\ket{\psi}}	& \qw 				& \qw 	&\qw 	& \multigate{1}{M_2}	& \ctrl{1} 	& \qw & \phi \\
	\lstick{\rho_m'}	& \multigate{1}{M_1^{+}}	& \ctrl{1}	& \gate{H}	& \ghost{M_2}			& \targ	& \qw &\\
	\lstick{\rho_m'}	& \ghost{M_1^{+}}		& \targ	& \qw	& \qw 				& \qw 	& \qw & \\
	%\gategroup{1}{3}{3}{9}{.7em}{--}
}
\end{align*}
\caption{Gate sequence in order to apply the $\Lambda_{\pi /12}$ gate to an arbitrary input state $\ket{\psi}$~\cite{Bravyi}. The states $\rho_m$ are the magic states obtained from the magic state distillation protocol. $M_1$ and $M_2$ are measurements of the observable $Z \otimes Z$. In the case of $M_1^{+}$, one needs to post-select upon obtaining the ``+1" outcome. Depending of the result of the measurement $M_2$, either a  $\Lambda_{\pi /12}$ or $\Lambda_{-\pi /12}$ gate is applied and the output state is $\phi$. The remaining two qubits are discarded.}
\label{fig:universalgate}
\end{figure}
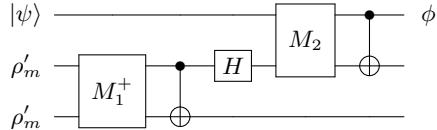

Due to the errors on the applied Clifford gates in the decoding procedure, distilling states arbitrarily close to the magic state is no longer possible, thus the distilled output state will have the form,
\begin{align}
\label{eq:noisyoutput}
\rho_{m}' = (1-\e') \ketbra{T_0}{T_0} + \e' \ketbra{T_1}{T_1},
\end{align}
and the value of the error $\e'$ will be fixed away from 0. However, the application of a universal non-Clifford gate is still possible with such a state, albeit with a certain level of noise. Following the procedure laid out in Ref.~\cite{Bravyi}, shown in Figure~\ref{fig:universalgate}, one can achieve the application of the gate $\Lambda_{\pi/12}$ to an arbitrary state $\ket{\psi} = a\ket{0} + b\ket{1}$ by using multiple copies of the noisy magic state $\rho_m'$. The fidelity of the applied gate is
\begin{align*}
F = 1 - \dfrac{12 \e' |a|^2 (1-|a|)^2}{ 3 + (1-2\e')^2} \ge 1-\e',
\end{align*}
where $\e'$ is the error rate of the distilled magic state given in Equation~\ref{eq:noisyoutput}. After multiple iterations of the distillation protocol, if the error of initial state was below the threshold for state distillation, the error rate $\e'$ will be approximately $p_1/2 + 13p_2/9$. Thus, in order to apply the universal gate with fidelity above $1-\e'$, one would typically have to apply a final fault-tolerant iteration in order to reduce the error rates of the original Clifford gates to a smaller logical error, as will be discussed in Section~\ref{sec:GateComparison}.

\subsection{Comparing faulty magic state distillation and fault-tolerant magic state distillation}
\label{sec:GateComparison}
Fault-tolerant quantum computation is a method to reduce the error rate of a given quantum operation using states and gates encoded into a higher-dimensional Hilbert space. Such encodings, combined with the ability of projective measurement and post-selection, provide a means to increase the fidelity of our quantum gates at the expense of using additional qubits and quantum gates to perform the desired encodings of states and encoded operations~\cite{ShorFT, Gottesman97, KLZ, Preskill, Aharonov, AGP06}.

Consider a scenario where we are presented with a physical device that is not restricted in the number of qubits at its disposal but limited in the number of gates that can be applied coherently. We could then envision two different methods for applying the magic state distillation procedure, either by applying the protocol with the faulty gates at our disposal or first performing a fault-tolerant encoding of the qubits and then performing the distillation protocol with less noisy encoded logical quantum gates. The faulty distillation protocol, as described and analyzed in Section~\ref{sec:depolarizing_noise}, suffers from a reduced convergence rate and a fundamental limit in the ability to distill a magic state with high fidelity. Thus using a fault-tolerance encoding of the Clifford gates appears to be advantageous. However the cost associated with encoding each of the Clifford gates in a fault-tolerant encoding outweighs the savings one would obtain in the number of Clifford gates one would need to apply at lower noise rates in the faulty magic state distillation scheme. 

\begin{figure}[h]
\includegraphics[width=0.66\textwidth]{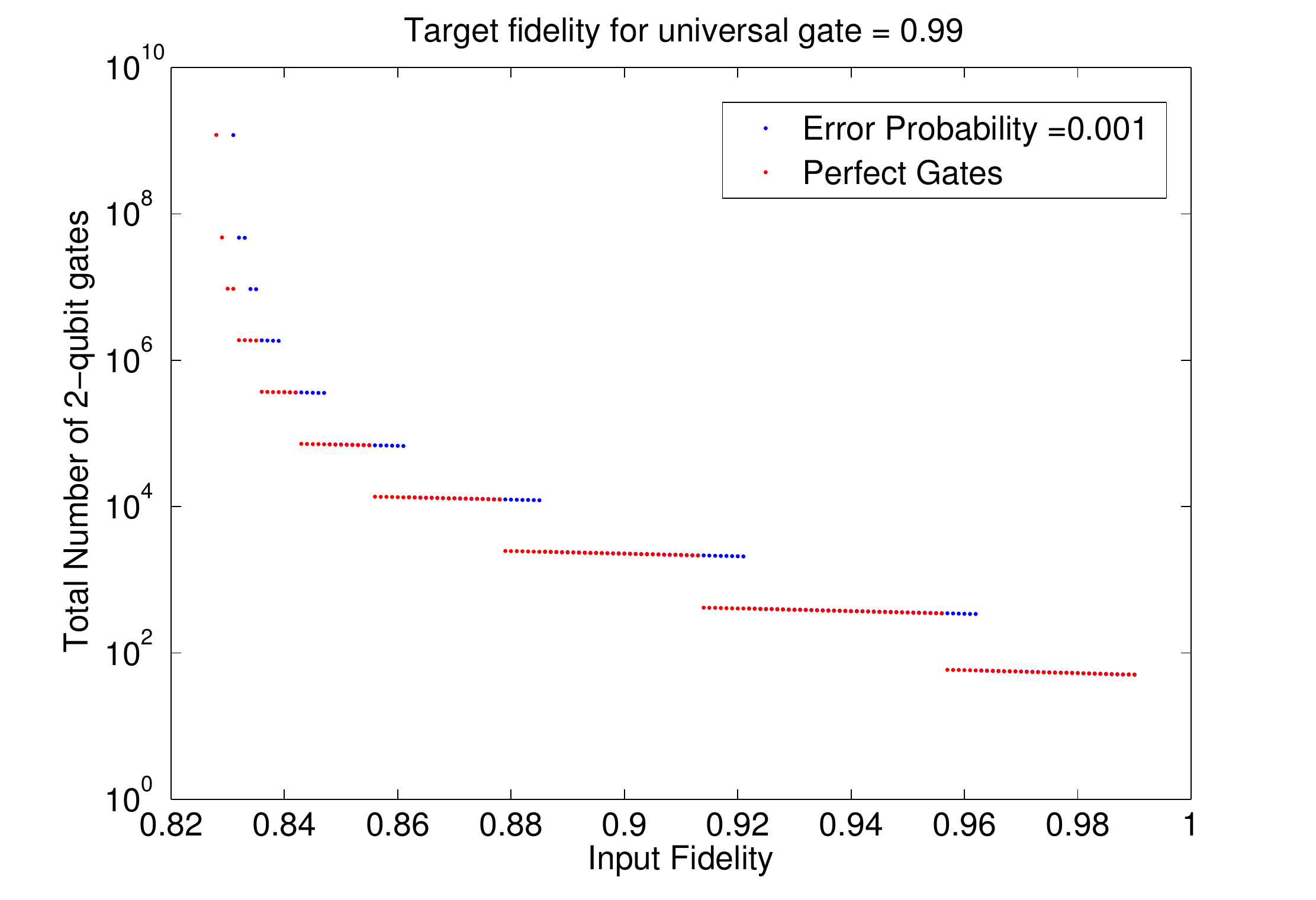}
\caption{Total number of two-qubit gates as a function of input fidelity for the state distillation scheme to achieve an output fidelity of 0.99 for the magic state. The red dots denote the number of gates required for perfect gates and can be thought of as a lower bound in the number of fault-tolerant encoded logical gates that would be required.  The blue dots represent the total number of gates for the faulty distillation protocol, where the gate error probability $E_1$, $E_2$ is 0.001. Note that the $y$-axis in this figure is plotted logarithmically in order to capture that the increase in the number of gates for each iteration of the distillation scheme scales approximately by a factor of 5.}
\label{fig:GateComparison}
\end{figure}

Figure~\ref{fig:GateComparison} plots the total number of two-qubit gates that would be required for either scheme to achieve a desired final fidelity for the magic state. The red dots plotted in Figure~\ref{fig:GateComparison} are the number of gates in the ideal setting with no noise, this can be thought of as a lower bound of the number of \emph{logical} fault-tolerant gates that would need to be applied for any fault-tolerant encoding. As expected, the number of gates exhibit a step function behaviour, characteristic of the number of iterations of the distillation subroutine that would be required to achieve the final fidelity target. The smearing of the steps is due to the probability of measuring the trivial syndrome, which is reduced for initial states with lower fidelity. One notices that the number of iterations for the fault-tolerant magic state distillation is lower than that of the faulty procedure, this is due to the fact that noise in faulty Clifford gates can decrease the convergent rate for states along the magic state axis and will lower the probability of measuring the trivial syndrome. Thus if one were to be presented with faulty Clifford gates and were to encode the states such that the error rate of the encoded operations were negligible, then the total number of encoded \emph{logical} gates required would be lower than the number of faulty Clifford gates to increase the fidelity of the magic state to a desired level. However, in the regime of typical one and two-qubit gate errors, $10^{-3}$--$10^{-2}$, the fault-tolerant encodings whose threshold rates are above these levels typically use on the order of $100$--$1000$ two-qubit physical gates per encoded logical gate~\cite{Cross}. As such, in order for a fault-tolerant encoded scheme to show an improvement in the total number of gates required for the full distillation protocol, the encoded scheme would have to use on the order of $100$--$1000$ times less logical gates than that of the faulty distillation scheme. Consider the results from Figure~\ref{fig:GateComparison}, which compares the number of faulty gates, with error probability $10^{-3}$, and the ideal case of encoded gates with no error. The reduction in the number of gates with no error compared to the faulty distillation scheme is greatest at the jumps in the number of gates that occur due to an increased number of iterations required. As such, the faulty distillation scheme may require an additional iteration of the distillation scheme, which is just repeating 5 times the previous level of distillation, leading to an overall gate cost on the order of 5 times more gates. However, as mentioned above, in order to reduce the error rate of the encoded states to be below that of the original error rate, each encoded \emph{logical} gate would require on the order of 100--1000 \emph{physical} gates. Therefore, while the number of logical gates is lower than the number of faulty gates required, the cost of each of the logical gates would drive the overall number of gates required in a fault-tolerant scheme to be much larger. We thus conclude that using faulty Clifford gates is more efficient for magic state distillation at noise levels comparable to those in current quantum information implementations. This analysis considers the number of two-qubit gates that would be required in the distillation protocol for either scheme as typical numbers of two-qubit gates have been studied extensively in past works~\cite{Cross}, however such an analysis could be extended to one-qubit gates and the authors believe that the behaviour will be equivalent.

Finally, one should note that fault-tolerance would be required if the errors in the one and two-qubit gates were too high, preventing the distilled state to achieve the desired target fidelity for a prepared magic state after multiple rounds of the distillation. This is due to the error of the distilled state always being approximately bounded from below by $p_1/2 + 13p_2/9$, where $p_1$ and $p_2$ are the error probabilities of the one and two-qubit gates, respectively. Therefore, if one requires to reduce the error of the distilled magic state, and subsequently the applied universal gate using the magic states, to be below some small target threshold, one would need to reduce the error rates of the one and two-qubit gates through a method such as fault-tolerance. However, this step would only have to be applied in the final iterations of the protocol in order to get the distilled state over the fidelity hump posed by the errors in the Clifford gates at one's disposal. 

%%
%% SECTION
%%
\section{Conclusion}

We have both analytically and numerically analyzed the evolution of quantum states under the five-qubit magic state distillation protocol without the presence of the dephasing transformation for perfect Clifford gates. We have characterized which input states converge to the magic state under repeated application of the distillation subroutine. Section~\ref{sec:nodephasing} provides results showing that not all states on a plane with fidelity over the threshold converge to magic state after multiple iterations of subroutine without dephasing. However, some quantum states which are undistillable become distillable with the addition of the dephasing transformation for high enough fidelity, as was shown by Bravyi and Kitaev~\cite{Bravyi}. Therefore, the fact that some quantum states contribute to universal quantum computation is dependent on the ability to access the dephasing transformation. Thus, on a more fundamental level, one may ask what is the role of the dephasing transformation in universal quantum computation? Also noted in Section~\ref{sec:nodephasing}, there are states below the theoretical threshold derived for states along the magic state axis, which converge to the magic state after many repetitions of the distillation routine, and as such if one had access to such states off the magic state axis, one could lower the threshold for distillation to 0.825. An interesting direction for future research would be to investigate whether other distillation schemes~\cite{Ben, Ben2,Campbell1, Campbell2, Anwar} are also robust to such noise perturbations.

Additionally, we studied the effect of noisy Clifford gates on the output fidelity, and the error rate in the universal gate induced by noisy distillation of magic states. Fortunately, based on the average error strength given in a recent benchmarking experiment~\cite{Ryan_benchmarking}, the state after absorbing all the noise in the Clifford gates is not far away from the pure magic state. This means that magic state distillation is robust against the typical noise levels in current experimental implementations. Nevertheless, due to the errors on the applied Clifford gates in the distillation procedure, one cannot produce an output state arbitrarily close to the magic state. In order to further reduce the error on the magic state, and the subsequent universal gate application, one would need to introduce a fault-tolerant encoding of the Clifford gates to reduce their noise. However, it is only at this point that a fault-tolerant encoding would be beneficial. We showed in Section~\ref{sec:GateComparison} that although introducing a fault-tolerant encoding at the beginning of the distillation protocol may seem appealing due to the increased convergence rate of the protocol, it is highly inefficient due to the number of gates that are typically required in any fault-tolerant scheme to reduce the error rate of current implementations of quantum information processors~\cite{Cross}. Since fault-tolerance would not be required to perform magic state distillation, other than in the final iterations of the protocol in order to apply a universal gate with very high fidelity, this work confirms that multiple rounds of faulty magic state distillation is a realistic scheme for future small-scale quantum information processors. 

%%
%% SECTION
%%
\section{Acknowledgements}
\emph{Author Contributions} -- All authors contributed to the development and discussion of the results in this work. T.J.~and~Y.Y. co-wrote the manuscript.

T.J. acknowledges the support of NSERC through the Alexander~Graham~Bell~CGS--M scholarship. Y.Y~acknowledges the support of the China~Scholarship~Council. B.H.~acknowledges the support of NSERC through the NSERC Undergraduate Student Research Award (USRA). The authors would like to thank Joseph~Emerson and Ben~Reichardt for stimulating insightful discussion throughout this project. This work was additionally funded through QuantumWorks, CIFAR, and Industry Canada.

%%
%% BIBLIOGRAPHY
%%
\bibliographystyle{ieeetr}
\bibliography{bibtex_magic}

\end{document}